%% file: main.tex
\documentclass[runningheads]{llncs}

\usepackage[T1]{fontenc}
\usepackage{amssymb}
\usepackage{graphicx}
\usepackage[table,xcdraw]{xcolor}
\usepackage[pagebackref,breaklinks,colorlinks]{hyperref}

\begin{document}

\title{d-Sketch: Improving Visual Fidelity of Sketch-to-Image Translation with Pretrained Latent Diffusion Models without Retraining}

\titlerunning{d-Sketch}

\author{
Prasun Roy\inst{1}\orcidID{0000-0002-1733-5670} \and
Saumik Bhattacharya \inst{2}\orcidID{0000-0003-1273-7969} \and
Subhankar Ghosh\inst{1}\orcidID{0000-0003-3242-3406} \and
Umapada Pal\inst{3}\orcidID{0000-0002-5426-2618} \and
Michael Blumenstein\inst{1}\orcidID{0000-0002-9908-3744}
}

\authorrunning{P. Roy \emph{et al.}}

\institute{
University of Technology Sydney, NSW 2007, Australia\\
\email{\{prasun.roy, subhankar.ghosh\}@student.uts.edu.au}\\
\email{michael.blumenstein@uts.edu.au} \and
Indian Institute of Technology Kharagpur, WB 721302, India\\
\email{saumik@ece.iitkgp.ac.in} \and
Indian Statistical Institute Kolkata, WB 700108, India\\
\email{umapada@isical.ac.in}
}

\maketitle

\input{sec/0_abstract}
\input{sec/1_introduction}
\input{sec/2_related_work}
\input{sec/3_method}
\input{sec/4_experiments}
\input{sec/5_conclusions}

\bibliographystyle{splncs04}
\bibliography{main}

\end{document}

%% file: sec/0_abstract.tex
\vspace{-0.8em}

\begin{abstract}
Structural guidance in an image-to-image translation allows intricate control over the shapes of synthesized images. Generating high-quality realistic images from user-specified rough hand-drawn sketches is one such task that aims to impose a structural constraint on the conditional generation process. While the premise is intriguing for numerous use cases of content creation and academic research, the problem becomes fundamentally challenging due to substantial ambiguities in freehand sketches. Furthermore, balancing the trade-off between shape consistency and realistic generation contributes to additional complexity in the process. Existing approaches based on Generative Adversarial Networks (GANs) generally utilize conditional GANs or GAN inversions, often requiring application-specific data and optimization objectives. The recent introduction of Denoising Diffusion Probabilistic Models (DDPMs) achieves a generational leap for low-level visual attributes in general image synthesis. However, directly retraining a large-scale diffusion model on a domain-specific subtask is often extremely difficult due to demanding computation costs and insufficient data. In this paper, we introduce a technique for sketch-to-image translation by exploiting the feature generalization capabilities of a large-scale diffusion model without retraining. In particular, we use a learnable lightweight mapping network to achieve latent feature translation from source to target domain. Experimental results demonstrate that the proposed method outperforms the existing techniques in qualitative and quantitative benchmarks, allowing high-resolution realistic image synthesis from rough hand-drawn sketches.

\keywords{Sketch-to-Image translation \and Latent diffusion models.}
\end{abstract}

%% file: sec/1_introduction.tex
\section{Introduction}
\label{sec:introduction}

\begin{figure}[h]
  \centering
  \includegraphics[width=\linewidth]{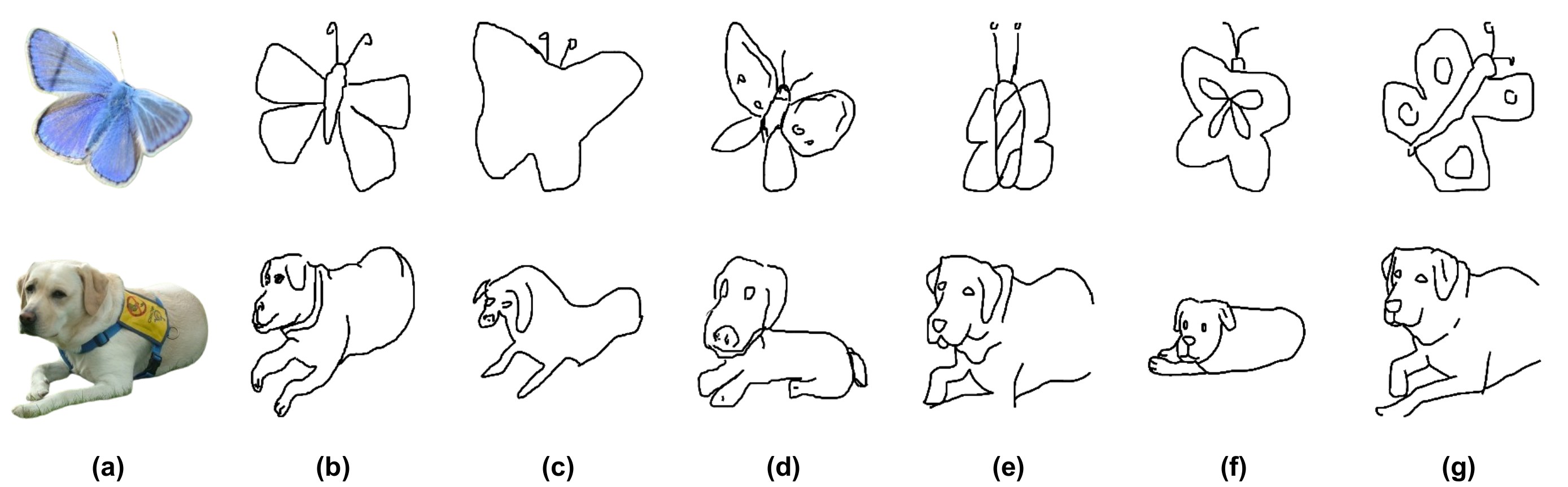}
  \caption{Structural ambiguity in hand-drawn sketches. \textbf{(a)} Subject image. \textbf{(b)--(g)} Freehand sketches drawn by different users. The examples are from the Sketchy dataset \cite{sangkloy2016sketchy}.}
  \label{fig:sketch_ambiguity}
\end{figure}

Freehand sketches provide simple and intuitive visual representations of natural images, allowing humans to understand and envision complex objects with a few sparse strokes. The convenience of modifying such minimalistic stroke-based representations to conceptualize semantic image manipulation is one key motivation for researchers to explore sketch-to-image translation. There are two primary objectives for such conditional generation -- the synthesized image should be \emph{visually realistic} and \emph{structurally consistent} with the input sketch, enabling perceptually appealing image generation from hand-drawn sketches irrespective of the artistic expertise of users. However, the intriguing premise becomes substantially challenging due to the practically unavoidable ambiguities in freehand sketches. For example, sketches of a specific object drawn by different persons can widely differ in stroke density and structural adherence depending on artistic abilities, as illustrated in Fig. \ref{fig:sketch_ambiguity} with samples from the Sketchy dataset \cite{sangkloy2016sketchy}.


\noindent
Consequently, this problem forces the generative algorithms to balance the trade-off between visual realism and intended shape. Existing GAN-based \cite{goodfellow2014generative,mirza2014conditional} methods primarily address sketch-to-image translation in two ways -- a direct mapping between domains with conditional GANs \cite{chen2020deepfacedrawing,chen2018sketchygan,gao2020sketchycoco,ghosh2019interactive,isola2017image,li2019linestofacephoto,li2020deepfacepencil,lu2018image,wang2018high,zhu2017unpaired} or modification in latent space using GAN inversions \cite{an2023sketchinverter,zhu2016generative}. However, such techniques often require application-specific data, optimization objectives, and complex learning strategies but occasionally fail to produce stable outcomes. Additionally, these methods operate on limited sets of task-specific object classes, resulting in poor generalization for unseen categories.

\noindent
More recently, denoising diffusion probabilistic models \cite{dhariwal2021diffusion,ho2020denoising,nichol2021improved,sohl2015deep} have demonstrated unprecedented improvements in the perceptual quality of general image synthesis. With sufficiently large annotated datasets \cite{schuhmann2022laion,schuhmann2021laion}, text-conditioned diffusion models \cite{ramesh2022hierarchical,ramesh2021zero,rombach2022high,saharia2022photorealistic} have achieved state-of-the-art results across multiple vision tasks, such as image generation, super-resolution, and inpainting. However, due to high structural ambiguities in hand-drawn sketches and the lack of sufficient paired sketch-image data, large-scale diffusion models have seen limited success in sketch-to-image translation. Furthermore, training such architectures from scratch is often computationally demanding and heavily infrastructure-dependent, which limits the scope of adopting the rich generative capabilities of latent diffusion models into sketch-to-image translation.

\noindent
In this paper, we propose a novel method for photorealistic image generation from freehand sketches leveraging the learned feature space of a pre-trained latent diffusion model \cite{rombach2022high}. We achieve this by introducing a learnable lightweight feature mapping network to perform latent code translation between source (\emph{sketch}) and target (\emph{image}) domains. The proposed approach provides a more stable optimization than GANs without requiring to train the latent diffusion model, thus mitigating the instability of GANs and the high computational overhead of large-scale diffusion models. Furthermore, unlike the existing methods, the proposed technique generalizes well beyond task-specific data distribution, significantly improving the generative performance on unseen object categories.

\vspace{1.0em}

\noindent
\textbf{Contributions:} The main contributions of the proposed work are as follows.

\begin{enumerate}
  \item We introduce an efficient method of photorealistic image generation from freehand sketches by providing structural guidance to a pre-trained latent diffusion model without retraining.
  \item The proposed approach achieves significantly better generalization beyond the observed data distribution, outperforming existing task-specific methods.
\end{enumerate}

\noindent
The remainder of the paper is organized as follows. Sec. \ref{sec:related_work} provides a brief overview of existing sketch-to-image translation techniques. Sec. \ref{sec:method} discusses the background and technical details of the proposed method, followed by the experimental analyses in Sec. \ref{sec:experiments}. We conclude the paper by summarizing our findings and discussing the potential scopes in Sec. \ref{sec:conclusions}.

%% file: sec/2_related_work.tex
\section{Related Work}
\label{sec:related_work}

\textbf{Conditional GANs:} Image-to-Image translation using conditional GANs is a widely explored method of directly transforming freehand sketches into images. Early architectural improvements introduced a Markovian discriminator \cite{isola2017image} for better retention of high-frequency correctness in paired image-to-image translation. A subsequent approach \cite{zhu2017unpaired} extended the idea to unpaired data by enforcing cycle consistency between source and target domains. In \cite{wang2018high}, the authors used coarse-to-fine generators, multi-scale discriminators, and an additional feature-matching loss for generating higher-resolution images. In \cite{park2019semantic}, the authors achieved generational improvements in semantic image manipulation by introducing spatially-adaptive normalization. The initial work exclusively on multi-class sketch-to-image translation proposed a masked residual unit \cite{chen2018sketchygan}, accommodating fifty object categories. Another approach proposed a contextual GAN \cite{lu2018image} to learn the joint distribution of the sketch and corresponding image. Researchers also explored interactive generation \cite{ghosh2019interactive} using a gating mechanism to suggest the probable completion of a partial sketch, followed by rendering the final image with a pre-trained image-to-image translation model \cite{wang2018high}. In \cite{gao2020sketchycoco}, the authors proposed a multi-stage class-conditioned approach for object-level and scene-level image synthesis from freehand sketches, improving the perceptual baseline over direct generations \cite{isola2017image}, contextual networks \cite{lu2018image}, and methods based on scene graphs \cite{ashual2019specifying,johnson2018image} or layouts \cite{zhao2019image}. In \cite{wang2022unsupervised}, the authors achieved similar goals with an unsupervised approach by introducing a standardization module and disentangled representation learning.

\vspace{1.0em}

\noindent
\textbf{GAN inversions:} The main objective of GAN inversion is to find a latent embedding of an image such that the original image can be faithfully reconstructed from the latent code using a pre-trained generator. Existing strategies for such inversions can be learning-based \cite{an2023sketchinverter,bau2019inverting,perarnau2016invertible,zhu2016generative}, optimization-based \cite{abdal2019image2stylegan,abdal2020image2stylegan++,creswell2018inverting,lipton2017precise,ma2018invertibility,ramesh2019spectral,voynov2020unsupervised}, or hybrid \cite{bau2019seeing,zhu2020domain}. In a learning-based inversion, an encoder learns to project an image into the latent space, minimizing reconstruction loss between the decoded (reconstructed) and original images. An optimization-based inversion estimates the latent code by directly solving an objective function. In a hybrid approach, an encoder first learns the latent projection, followed by an optimization strategy to refine the latent code. The rich statistical information captured by deep generative networks from large-scale data provides effective \emph{priors} for various downstream tasks, including sketch-to-image translation. In \cite{an2023sketchinverter}, the authors adopted a learning-based GAN inversion strategy using a multi-class deep generative network \cite{brock2018large}, pre-trained on the large-scale ImageNet dataset \cite{deng2009imagenet}, as \emph{prior} to achieve sketch-to-image translation for multiple categories. In \cite{xiang2022adversarial}, the authors introduced a framework for generalizing image synthesis to \emph{open-domain} object categories by jointly learning two \emph{in-domain} mappings (image-to-sketch and sketch-to-image) with \emph{random-mixed} strategy.

\vspace{1.0em}

\noindent
\textbf{Diffusion models:} A Denoising Diffusion Probabilistic Model (DDPM) \cite{ho2020denoising,sohl2015deep} is a parameterized Markov chain that learns to generate samples similar to the original data distribution after a finite time. In particular, DDPMs use variational inference to learn to iteratively reverse a stepwise \emph{diffusion} (noising) process. In \cite{song2020denoising}, the authors introduced Denoising Diffusion Implicit Models (DDIMs) by generalizing DDPMs using non-Markovian diffusion processes with the same learning objective, leading to a deterministic and faster generative process. Recent advances \cite{dhariwal2021diffusion,nichol2021improved} have shown that diffusion models can achieve generational improvements in the visual quality and sampling diversity over GANs while providing a more stable and straightforward optimization objective. The most prolific application of diffusion models in recent literature is text-conditioned image generation \cite{ramesh2022hierarchical,ramesh2021zero,rombach2022high,saharia2022photorealistic} and modification \cite{bar2022text2live,brooks2023instructpix2pix,hertz2023prompt,mokady2023null}, utilizing a pre-trained language-image model \cite{radford2021learning} to embed the conditioning prompt. In \cite{choi2021ilvr}, the authors guided the generative process with an iterative latent variable refinement to produce high-quality variations of a reference image. In \cite{ruiz2023dreambooth}, the authors introduced a class-specific prior preservation loss to finetune an existing text-to-image diffusion model for \emph{personalized} manipulation of a specific subject image from a few observations. Emerging alternative approaches also involved Stochastic Differential Equations (SDEs) to guide the generative process following score-based \cite{meng2021sdedit} or energy-based \cite{xing2023inversion,zhao2022egsde} objectives. More recent attempts for sketch-to-image translation involved multiple objectives \cite{wang2022diffsketching}, multi-dimensional control \cite{cheng2023adaptively}, or latent code optimization \cite{voynov2023sketch}. In \cite{wang2022diffsketching}, the authors used an additional network to reconstruct the input sketch from the generated image. The denoising process was optimized using a cumulative objective function consisting of the \emph{perceptual similarity} (between the input and reconstructed sketches) and \emph{cosine similarity} (between the input and generated images) measures. In \cite{cheng2023adaptively}, the authors provided three-dimensional controls over image synthesis from the strokes and sketches to manipulate the balance between \emph{perceptual realism} and \emph{structural faithfulness} during the conditional denoising process. In \cite{voynov2023sketch}, the authors introduced a lightweight mapping network for providing structural guidance to a pre-trained latent diffusion model \cite{rombach2022high}. While the method avoided training a dedicated diffusion network, the \emph{differential guidance} made sampling images computationally even more demanding than a large-scale model itself.

%% file: sec/3_method.tex
\section{Method}
\label{sec:method}

\subsection{Preliminaries}\label{sec:method_preliminaries}
\textbf{Diffusion models:} A Denoising Diffusion Probabilistic Model (DDPM) defines a Markov chain that learns to generate samples to match the input data distribution over a finite time. The process consists of \emph{forward diffusion} that iteratively perturbs an input by adding noise according to a scheduler, followed by \emph{backward denoising} that learns to reverse the mapping to recover the original input from noise. Given a data distribution $x_0 \sim q(x_0)$, the \emph{forward diffusion} defines an iterative noising process $q$ that adds Gaussian noise over $T$ finite steps, gradually perturbing the input sample $x_0$ to produce latents $\{x_1, ..., x_T\}$ as follows.

\begin{equation}\label{eq:eq1}
  q(x_1, ..., x_T | x_0) := \prod_{t=1}^{T} q(x_t | x_{t-1})
\end{equation}

\begin{equation}\label{eq:eq2}
  q(x_t | x_{t-1}) := \mathcal{N}(x_t; \sqrt{1-\beta_t} ~x_{t-1}, \beta_t \mathbf{I})
\end{equation}

\noindent
where $\beta_t \in (0, 1)$ denotes the variance of the Gaussian noise at time $t \sim [1, T]$. Rewriting Eq. \ref{eq:eq2} with $\alpha_t = 1 - \beta_t$ and $\overline{\alpha}_t = \prod_{i=1}^{t} \alpha_i$, Ho \emph{et al.} \cite{ho2020denoising} deduced a closed-form expression to sample an arbitrary step of the noising process, directly estimating $x_t$ from $x_0$ as the following marginal distribution.

\begin{equation}\label{eq:eq3}
  q(x_t | x_0) = \mathcal{N}(x_t; \sqrt{\overline{\alpha}_t} ~x_0, (1-\overline{\alpha}_t) \mathbf{I})
\end{equation}

\noindent
With sufficiently large $T$ and a well-defined schedule of $\beta_t$, the latent $x_T$ closely resembles a Gaussian distribution. If the reverse distribution $q(x_{t-1} | x_t)$ is known, sampling $x_T \sim \mathcal{N}(0, \mathbf{I})$ and iteratively running the process in reverse can yield a sample from $q(x_0)$. However, as $q(x_{t-1} | x_t)$ depends on the entire data distribution, the \emph{backward denoising} process can be approximated by a learnable network, parameterized with $\theta$, as follows.

\begin{equation}\label{eq:eq4}
  p_\theta(x_{t-1} | x_t) := \mathcal{N}(x_{t-1}; \mu_\theta(x_t, t), \Sigma_\theta(x_t, t))
\end{equation}

\noindent
Ho \emph{et al.} \cite{ho2020denoising} also observed that learning to predict the added noise $\epsilon \sim \mathcal{N}(0, \mathbf{I})$ worked best for estimating $x_0$ with the following formulation.

\begin{equation}\label{eq:eq5}
  x_0 = \frac{1}{\sqrt{\alpha_t}} \left( x_t - \frac{\beta_t}{\sqrt{1 - \overline{\alpha}_t}} ~\epsilon \right)
\end{equation}

\noindent
Most implementations adopt a U-Net architecture (parameterized with $\theta$) to predict the added noise, minimizing mean squared error as the learning objective.

\begin{equation}\label{eq:eq6}
  \mathcal{L}_{DM} = \mathbb{E}_{t \sim [1, T], ~x_0 \sim q(x_0), ~\epsilon \sim \mathcal{N}(0, \mathbf{I})} \bigg[ \| \epsilon - \epsilon_\theta(x_t, t) \|^2 \bigg]
\end{equation}

\subsection{Latent Code Translation Network (LCTN)}\label{sec:method_lctn}
\begin{figure}[h]
  \centering
  \includegraphics[width=\linewidth]{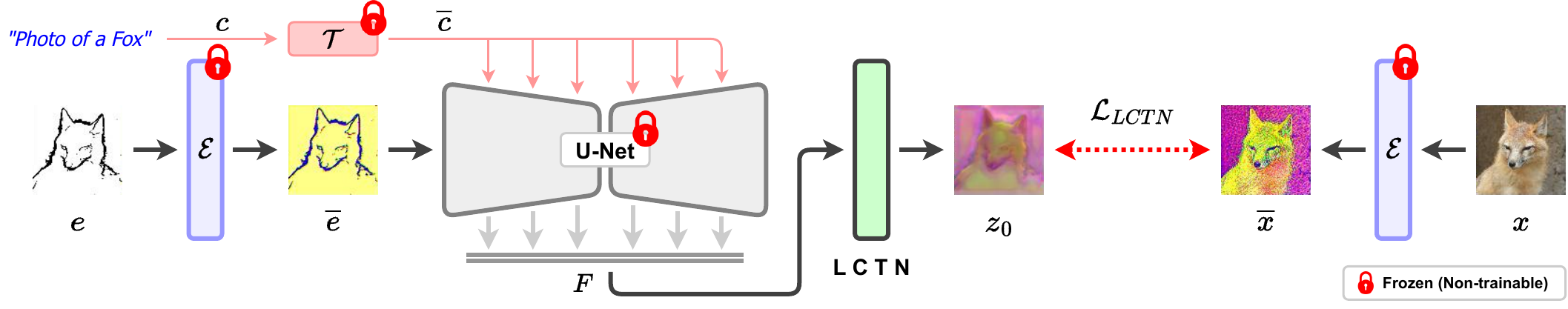}
  \caption{Proposed training strategy for the Latent Code Translation Network (LCTN).}
  \label{fig:lctn_training}
\end{figure}

We propose a learnable Latent Code Translation Network (LCTN) to shift the input latent space toward the target domain by exploiting the learned feature representations of a pre-trained Latent Diffusion Model (LDM) \cite{rombach2022high}. LCTN is trained with edge maps \cite{su2021pixel} instead of hand-drawn sketches to mitigate the structural ambiguities that arise from freehand sketches. Our experiments show that LCTN trained on edge maps works appreciably well during inference with freehand sketches. Given an image $x$, corresponding edge map $e$, and object class name $c$, we use the pre-trained image encoder $\mathcal{E}$ and text encoder $\mathcal{T}$ of LDM to compute the initial latent codes as, $\overline{x} = \mathcal{E}(x)$, $\overline{e} = \mathcal{E}(e)$, and $\overline{c} = \mathcal{T}(c)$. The input feature space $F$ is computed from the intermediate activation maps of LDM U-Net $\epsilon_\theta$, rescaled to have the same spatial dimensions, with a single denoising pass of $\overline{e}$ at timestep $t = 0$ using $\overline{c}$ as conditioning, $F = f_{\epsilon_\theta}(\overline{e}, \overline{c}, t)$. LCTN learns to project $F$ into the target latent code $z_0$ by minimizing the mean squared error, $\mathcal{L}_{LCTN} = \|z_0 - \overline{x}\|^2$. Architecturally, LCTN consists of a sequence of fully connected (FC) layers with 512, 256, 128, and 64 nodes, with each FC layer followed by ReLU activation and batch normalization. We illustrate the proposed training strategy for LCTN in Fig. \ref{fig:lctn_training}.

\begin{figure}[t]
  \centering
  \includegraphics[width=\linewidth]{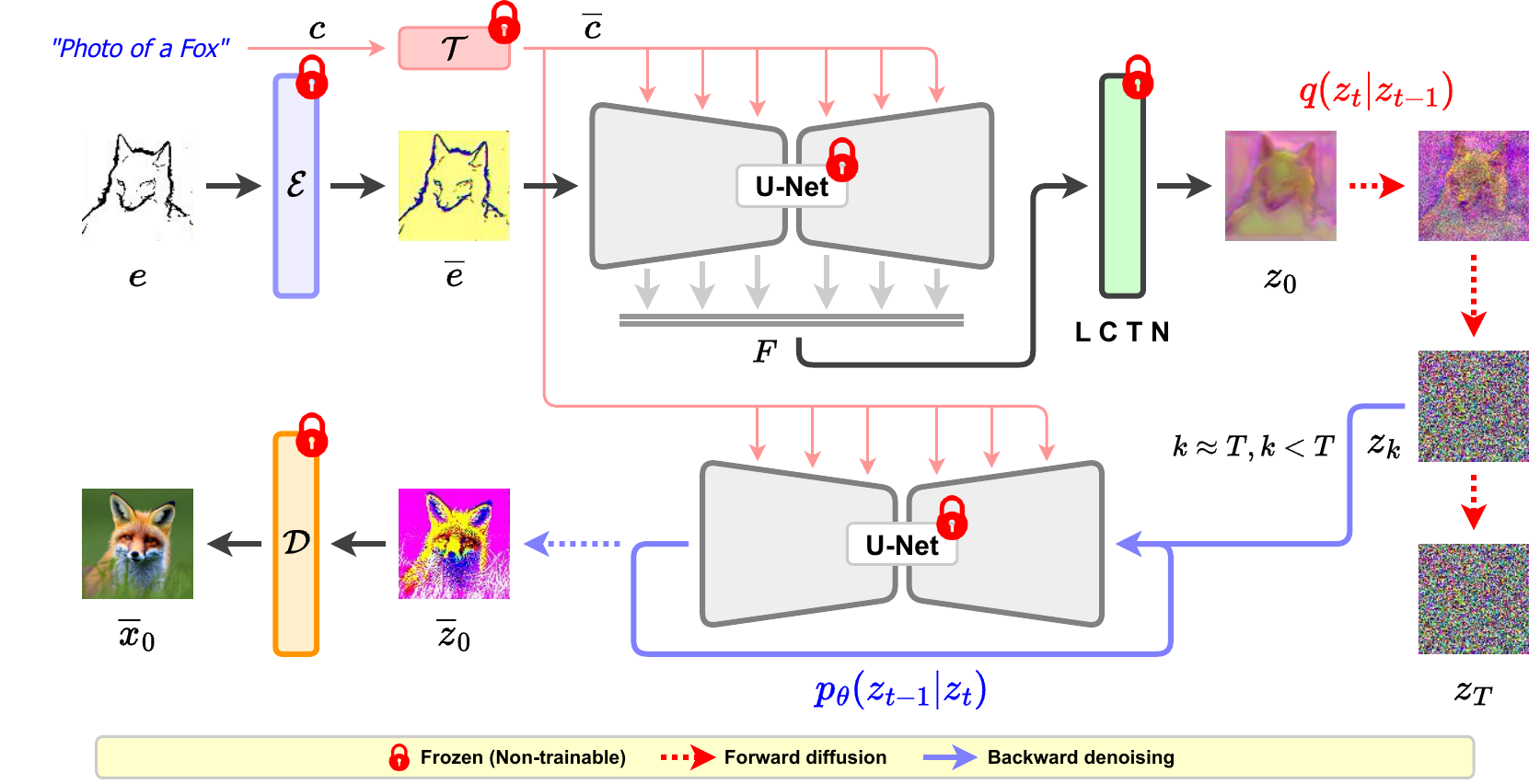}
  \caption{Proposed sampling strategy for the Latent Code Translation Network (LCTN).}
  \label{fig:lctn_sampling}
\end{figure}

\noindent
Ideally, if the domain translation by LCTN is accurate, we can readily decode $z_0$ into a high-quality photorealistic image using the pre-trained LDM decoder $\mathcal{D}$. However, due to high sparsity in the input edge maps (or sketches), LCTN-projected latent code lacks sufficient subtlety, leading to unrealistic images from direct decoding. We address the issue by first perturbing $z_0$ to $z_k$ over $k \sim [1, T]$ steps, where $1 < k < T$, followed by $T$ denoising iterations to get $\overline{z}_0$ from $z_k$. With a sufficiently large value of $k$, $z_k$ is close to an isotropic Gaussian distribution, $z_k \approx z_T \sim \mathcal{N}(0, \mathbf{I})$. However, strictly enforcing $k < T$ ensures minimal structural elements are retained in $z_k$. We observe that starting the backward denoising from $z_k$ instead of $z_T$ as the initial latent, followed by decoding the final latent $\overline{z}_0$, can produce photorealistic images while retaining the intended structural resemblance with the input edge map (or sketch). In our experiments, $0.7 \leqslant \frac{k}{T} \leqslant 0.9$ works best for most cases. We illustrate the proposed sampling strategy for LCTN in Fig. \ref{fig:lctn_sampling}.

%% file: sec/4_experiments.tex
\section{Experiments}
\label{sec:experiments}

\begin{figure}[t]
  \centering
  \includegraphics[width=\linewidth]{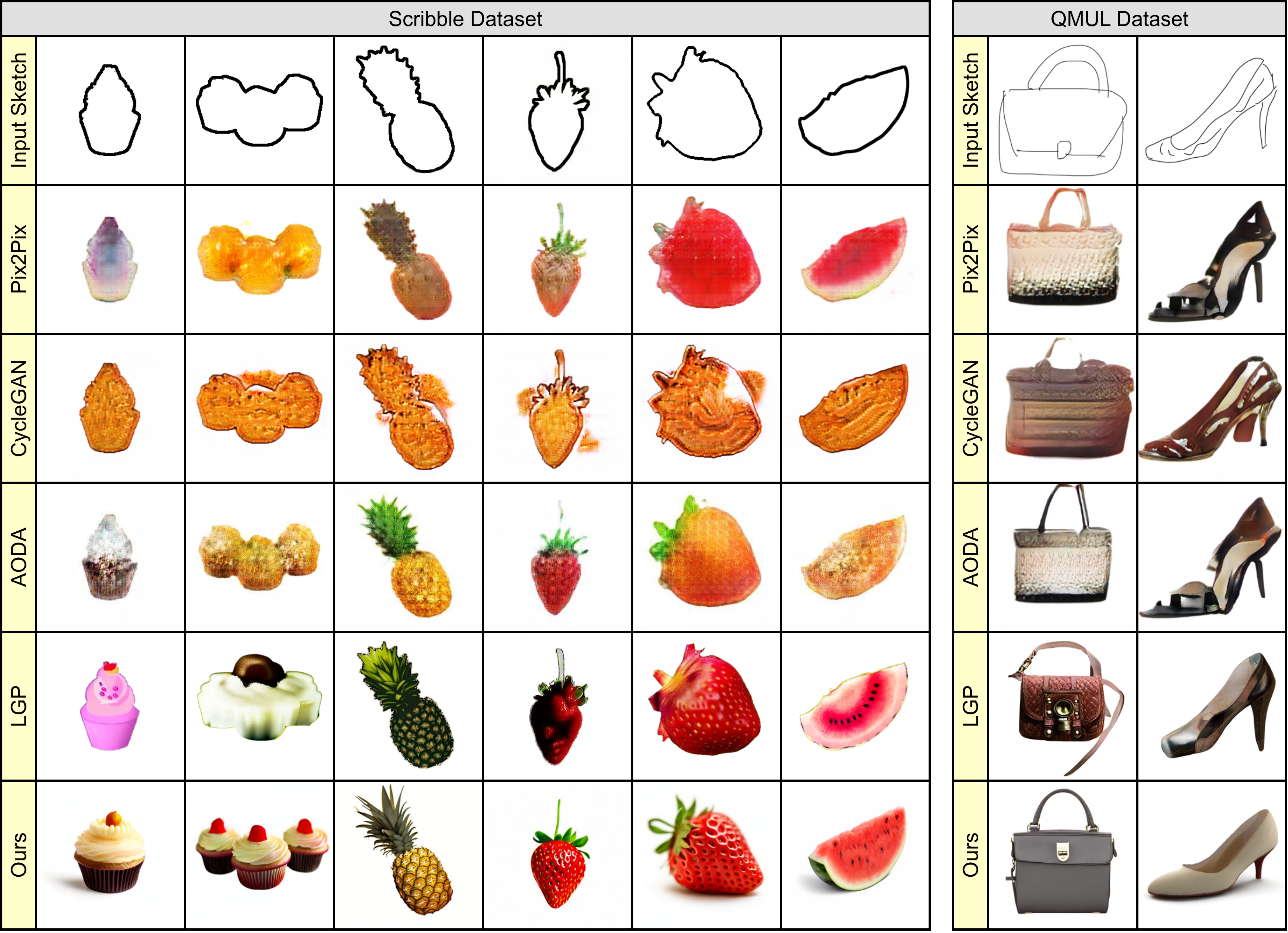}
  \caption{Qualitative comparison of the proposed method with existing sketch-to-image translation techniques -- Pix2Pix \cite{isola2017image}, CycleGAN \cite{zhu2017unpaired}, AODA \cite{xiang2022adversarial}, and LGP \cite{voynov2023sketch} on Scribble \cite{ghosh2019interactive} and QMUL \cite{song2017deep,yu2016sketch} datasets.}
  \label{fig:comparison}
  \vspace{-0.5em}
\end{figure}

\begin{figure}[t]
  \centering
  \includegraphics[width=\linewidth]{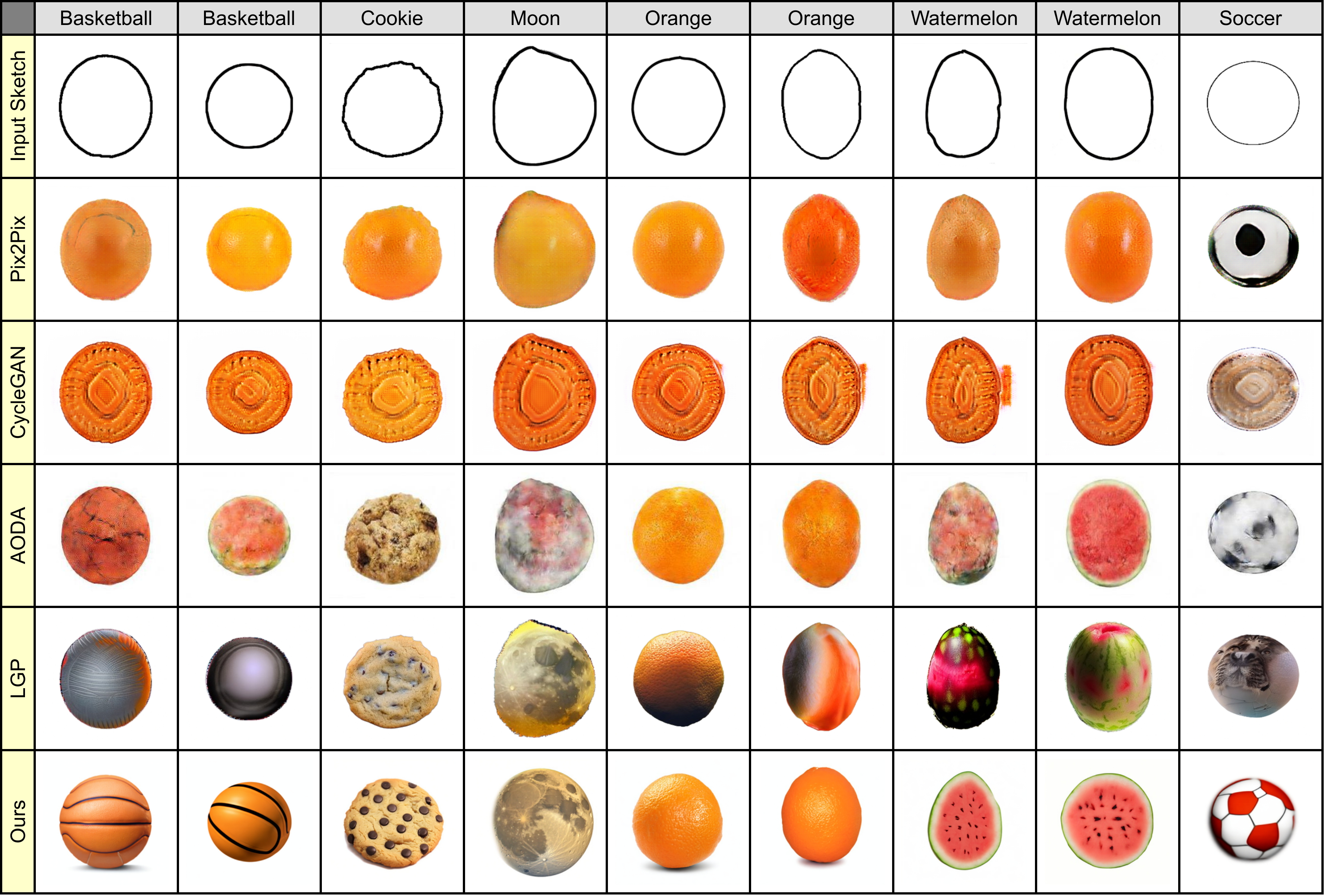}
  \caption{Qualitative comparison for distinct object classes with nearly identical shapes. The proposed method can produce high-quality, visually distinguishable objects in contrast to the ambiguous results generated by existing sketch-to-image translation techniques -- Pix2Pix \cite{isola2017image}, CycleGAN \cite{zhu2017unpaired}, AODA \cite{xiang2022adversarial}, and LGP \cite{voynov2023sketch}.}
  \label{fig:comparison_similar_shapes}
  \vspace{-0.5em}
\end{figure}

\noindent
\textbf{Datasets:} We evaluate the performance of the proposed method against existing sketch-to-image translation techniques \cite{isola2017image,voynov2023sketch,xiang2022adversarial,zhu2017unpaired} on three following datasets.

\vspace{0.5em}

\noindent
\textbf{(a) Scribble:} The Scribble dataset \cite{ghosh2019interactive} contains $256 \times 256$ image-sketch pairs of ten object classes (basketball, chicken, cookie, cupcake, moon, orange, pineapple, soccer, strawberry, and watermelon) having uniform white backgrounds. While the images in the dataset do not feature complex backgrounds, 60\% of the object classes share nearly identical circular shapes, which introduces significant ambiguities to the generative algorithms. We use 1512 image-sketch pairs \cite{xiang2022adversarial} (1412 train + 100 test) to train and evaluate all competing methods.

\vspace{0.5em}

\noindent
\textbf{(b) QMUL:} The QMUL dataset is a compilation \cite{xiang2022adversarial} of image-sketch pairs from three object categories -- shoe \cite{yu2016sketch}, chair \cite{yu2016sketch}, and handbag \cite{song2017deep} with uniform white backgrounds. Due to the structural ambiguities in the provided hand-drawn sketches, the dataset poses a substantial challenge to the generative algorithms. Following \cite{xiang2022adversarial}, we use 7850 freehand sketches of 3004 images for training and 691 freehand sketches of 480 images for evaluation.

\vspace{0.5em}

\noindent
\textbf{(c) Flickr20:} While Scribble \cite{ghosh2019interactive} and QMUL \cite{song2017deep,yu2016sketch} datasets provide significant structural challenges to the learning algorithms, the images do not contain perceptual complexities of natural backgrounds. To investigate the generative performances in such cases, we introduce a new dataset by collecting 10K (9500 train + 500 test) high-resolution images from \href{https://www.flickr.com/}{Flickr}, equally distributed over 20 animal classes -- bird, cat, cow, deer, dog, dolphin, elephant, fox, frog, giraffe, goat, horse, lion, monkey, pig, polar bear, rabbit, sheep, tiger, and zebra. The edge maps for these images are estimated with a pre-trained edge detector \cite{su2021pixel}.

\vspace{0.5em}

\noindent
\textbf{Implementation and experimental details:} The LCTN architecture consists of a sequence of four fully connected (FC) hidden layers having 512, 256, 128, and 64 nodes, with each FC layer followed by ReLU activation and batch normalization. A final FC layer projects the last hidden layer output to a 4D latent vector, representing a single spatial position in the 4-channel latent space $z_0$. We use the \emph{Stable Diffusion v2.1} (SD2.1) distribution for pre-trained text encoder, VAE and U-Net. LCTN is trained for 50000 iterations at a constant learning rate of 0.001 with 100 initial warm up steps on a single NVIDIA Quadro RTX 6000 GPU with a batch size of 4 and FP16 mixed precision. We keep the default image size of SD2.1 ($768 \times 768$) throughout all our experiments. LCTN is initialized with a normal distribution $\mathcal{N}(0, 0.02)$. We optimize the parameters of LCTN using stochastic Adam optimizer \cite{kingma2015adam} having $\beta$-coefficients (0.9, 0.999). For reproducibility, the code is officially available at \url{https://github.com/prasunroy/dsketch}. We have included the full-resolution visual results in the \textbf{\emph{supplementary material}}.

\vspace{0.5em}

\noindent
\textbf{Visual analysis:} For analyzing the perceptual quality of the generated images by our method, we perform a visual comparison with existing GAN-based \cite{isola2017image,xiang2022adversarial,zhu2017unpaired} and diffusion-based \cite{voynov2023sketch} sketch-to-image translation techniques. Fig. \ref{fig:comparison} demonstrates a qualitative comparison of the proposed method against Pix2Pix \cite{isola2017image}, CycleGAN \cite{zhu2017unpaired}, AODA \cite{xiang2022adversarial}, and LGP \cite{voynov2023sketch} on Scribble \cite{ghosh2019interactive} and QMUL \cite{song2017deep,yu2016sketch} datasets. Our method can generate highly detailed and perceptually appealing samples that are visibly superior to existing approaches while maintaining the intended structural resemblance with the input sketches.

\vspace{0.5em}

\noindent
\textbf{Visual analysis on ambiguous classes:} Occasionally, multiple visually distinguishable objects can have identical shapes. For example, 60\% object classes in the Scribble dataset \cite{ghosh2019interactive} have an identical circular structure (basketball, cookie, moon, orange, soccer, and watermelon), leading to nearly indistinguishable sketches for visibly distinguishable object categories. Therefore, it poses a substantial challenge to the generative algorithms for producing class-conditioned distinctive visual features in such ambiguous cases. Fig. \ref{fig:comparison_similar_shapes} shows a qualitative comparison of the proposed method against existing approaches \cite{isola2017image,voynov2023sketch,xiang2022adversarial,zhu2017unpaired} on ambiguous classes from the Scribble dataset \cite{ghosh2019interactive}. Pix2Pix \cite{isola2017image} and CycleGAN \cite{zhu2017unpaired} mostly fail to produce distinguishable objects. AODA \cite{xiang2022adversarial} and LGP \cite{voynov2023sketch} achieve limited success in producing photorealistic results. In contrast, our method can generate high-quality and visibly distinctive images with class-specific visual attributes of intended objects from virtually identical sketches.

\vspace{-0.5em}

\begin{table}[h]
\centering
\caption{Quantitative analysis of the proposed method on Scribble \cite{ghosh2019interactive} dataset.}
\label{tab:comparison_scribble}
\resizebox{\columnwidth}{!}{%
\begin{tabular}{lccccccc}
\hline
\rowcolor[HTML]{e0e0e0}
\textbf{Method} &
  ~\textbf{FID $\downarrow$}~ &
  ~~~~\textbf{IS $\uparrow$}~~~~ &
  ~\textbf{PSNR $\uparrow$}~ &
  ~\textbf{SSIM $\uparrow$}~ &
  ~\textbf{LPIPS $\downarrow$}~ &
  ~\textbf{ACC $\uparrow$}~ &
  ~\textbf{\textcolor{blue}{MOS $\uparrow$}}~ \\ \hline
Pix2Pix \cite{isola2017image}    & 333.1872 & 3.8027 & 13.3208 & 0.6082 & 0.3635 & 0.24 & 0.02 \\
CycleGAN \cite{zhu2017unpaired}  & 322.6855 & 3.6737 & 13.4177 & 0.5804 & 0.3003 & 0.33 & 0.01 \\
AODA \cite{xiang2022adversarial} & 353.9626 & 4.0133 & 12.4880 & 0.5588 & 0.3761 & 0.19 & 0.01 \\
LGP \cite{voynov2023sketch}      & 207.8677 & 8.4247 & ~5.6862 & 0.3171 & 0.5667 & 0.72 & 0.24 \\ \hline
Ours &
  \textbf{163.8978} &
  \textbf{9.9132} &
  \textbf{13.8737} &
  \textbf{0.6406} &
  \textbf{0.2839} &
  \textbf{0.75} &
  \textbf{0.72} \\ \hline
\end{tabular}%
}
\end{table}

\vspace{-2.5em}

\begin{table}[h]
\centering
\caption{Quantitative analysis of the proposed method on QMUL \cite{song2017deep,yu2016sketch} dataset.}
\label{tab:comparison_qmul}
\resizebox{\columnwidth}{!}{%
\begin{tabular}{lccccccc}
\hline
\rowcolor[HTML]{e0e0e0}
\textbf{Method} &
  ~\textbf{FID $\downarrow$}~ &
  ~~~~\textbf{IS $\uparrow$}~~~~ &
  ~\textbf{PSNR $\uparrow$}~ &
  ~\textbf{SSIM $\uparrow$}~ &
  ~\textbf{LPIPS $\downarrow$}~ &
  ~\textbf{ACC $\uparrow$}~ &
  ~\textbf{\textcolor{blue}{MOS $\uparrow$}}~ \\ \hline
Pix2Pix \cite{isola2017image}    & 189.7064 & \textbf{5.3261} & ~9.2383 & 0.5328 & 0.4013 & 0.6151 & 0.04 \\
CycleGAN \cite{zhu2017unpaired}  & 146.3326 & 5.1030          & ~9.5792 & 0.6050 & 0.3198 & 0.4486 & 0.01 \\
AODA \cite{xiang2022adversarial} & 216.7982 & 5.0196          & ~9.8943 & 0.5784 & 0.4152 & 0.6208 & 0.01 \\
LGP \cite{voynov2023sketch}      & 108.1720 & 5.1159          & ~5.4842 & 0.1710 & 0.6943 & 0.8770 & 0.35 \\ \hline
Ours &
  \textbf{~63.9208} &
  4.3687 &
  \textbf{11.8780} &
  \textbf{0.6677} &
  \textbf{0.3126} &
  \textbf{0.9899} &
  \textbf{0.59} \\ \hline
\end{tabular}%
}
\end{table}

\vspace{-2.5em}

\begin{table}[h]
\centering
\caption{Quantitative analysis of the proposed method on Flickr20 dataset.}
\label{tab:comparison_flickr20}
\resizebox{\columnwidth}{!}{%
\begin{tabular}{lccccccc}
\hline
\rowcolor[HTML]{e0e0e0}
\textbf{Method} &
  ~\textbf{FID $\downarrow$}~ &
  ~~~~\textbf{IS $\uparrow$}~~~~ &
  ~\textbf{PSNR $\uparrow$}~ &
  ~\textbf{SSIM $\uparrow$}~ &
  ~\textbf{LPIPS $\downarrow$}~ &
  ~\textbf{ACC $\uparrow$}~ &
  ~\textbf{\textcolor{blue}{MOS $\uparrow$}}~ \\ \hline
Pix2Pix \cite{isola2017image}    & 122.4473 & ~8.5337 & 10.1246 & 0.1553 & 0.7136 & 0.430 & 0.02 \\
CycleGAN \cite{zhu2017unpaired}  & 162.6837 & ~6.6324 & 10.6105 & 0.1261 & 0.7848 & 0.242 & 0.00 \\
AODA \cite{xiang2022adversarial} & 150.0852 & ~7.4056 & 10.0145 & 0.1478 & 0.7325 & 0.332 & 0.01 \\
LGP \cite{voynov2023sketch}      & ~81.4195 & 14.9779 & ~9.3839 & 0.1109 & 0.7553 & 0.794 & 0.42 \\ \hline
Ours &
  \textbf{~72.5475} &
  \textbf{15.7383} &
  \textbf{10.9339} &
  \textbf{0.2113} &
  \textbf{0.6811} &
  \textbf{0.876} &
  \textbf{0.55} \\ \hline
\end{tabular}%
}
\end{table}

\vspace{-0.5em}

\noindent
\textbf{Evaluation metrics:} We measure seven metrics to quantitatively evaluate the perceptual quality, structural consistency, and class accuracy in the generated images. Fr\'{e}chet Inception Distance (\textbf{FID}) measures the feature space similarity between real and generated images. Inception Score (\textbf{IS}) estimates the Kullback-Leibler (KL) divergence between the label and marginal distributions to measure the visual quality and class diversity of generated images. Peak Signal-to-Noise Ratio (\textbf{PSNR}) assesses the quality of generated images by estimating the deviation from real images. Structural Similarity Index Measure (\textbf{SSIM}) estimates the structural consistency in the generated images against the ground truth by considering image degradation as the perceived change in structural information. Learned Perceptual Image Patch Similarity (\textbf{LPIPS}) quantifies the perceptual similarity between real and generated images using the spatial feature maps obtained from a pre-trained deep convolutional network such as SqueezeNet in our experiments. We also estimate the classification accuracy (\textbf{ACC}) using a multi-class image classifier to measure the correctness of intended object classes in generated samples.

\vspace{0.5em}

\noindent
\textbf{Human evaluation:} Although the said metrics are widely used in the literature, perceptual quality assessment is an open challenge in computer vision. Therefore, we conducted an opinion-based user assessment among 45 individuals, where the volunteers were asked to select the most visually realistic sample that had the closest resemblance to a given sketch from a pool of images generated by the competing methods. The Mean Opinion Score (\textbf{\textcolor{blue}{MOS}}) is the average fraction of times a method received user preference over other methods. Tables \ref{tab:comparison_scribble}, \ref{tab:comparison_qmul}, and \ref{tab:comparison_flickr20} summarize the evaluation scores of different methods on the Scribble \cite{ghosh2019interactive}, QMUL \cite{song2017deep,yu2016sketch}, and Flickr20 datasets, respectively. In most cases, the proposed method achieves a better score than the existing sketch-to-image translation techniques \cite{isola2017image,voynov2023sketch,xiang2022adversarial,zhu2017unpaired} across different datasets, indicating superior perceptual quality, structural consistency, and class accuracy in the generated images.

\vspace{0.5em}

\begin{figure}[h]
  \centering
  \includegraphics[width=\linewidth]{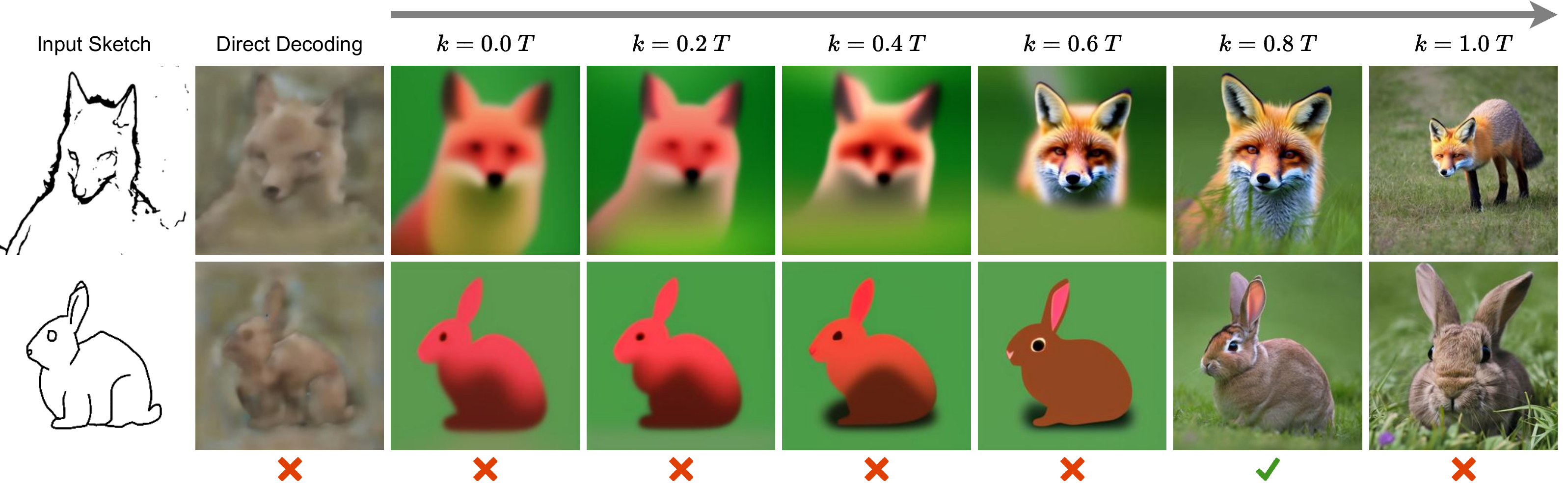}
  \caption{Visual analysis of balancing the trade-off between structural consistency and perceptual quality by selecting an optimal value of $k$ on the proposed Flickr20 dataset.}
  \label{fig:noising_scale}
\end{figure}

\noindent
\textbf{Analyzing the optimal value of \emph{k}:}
In the proposed method, $k \sim [1, T]$ is a crucial control parameter for balancing the trade-off between structural consistency and visual realism in the generated samples. As discussed in Sec. \ref{sec:method_lctn}, directly decoding the LCTN-projected latent $z_0$ through the image decoder $\mathcal{D}$ produces virtually unusable images $\mathcal{D}(z_0)$. For substantially lower values of $k$, the generated image $\overline{x}_0$ retains high structural accuracy but lacks photorealism. With increasing values of $k$, perceptual quality of $\overline{x}_0$ gradually improves at the expense of structural consistency. While the optimal value of $k$ varies among different datasets, $0.7 \leqslant \frac{k}{T} \leqslant 0.9$ works best for most cases in our experiments. Fig. \ref{fig:noising_scale} illustrates a visual analysis of balancing the trade-off between structural consistency and photorealism by selecting an optimal value of $k \approx T$, $k < T$.

\vspace{-0.5em}

\vspace{-1.0em}

\begin{figure}[t]
  \centering
  \includegraphics[width=\linewidth]{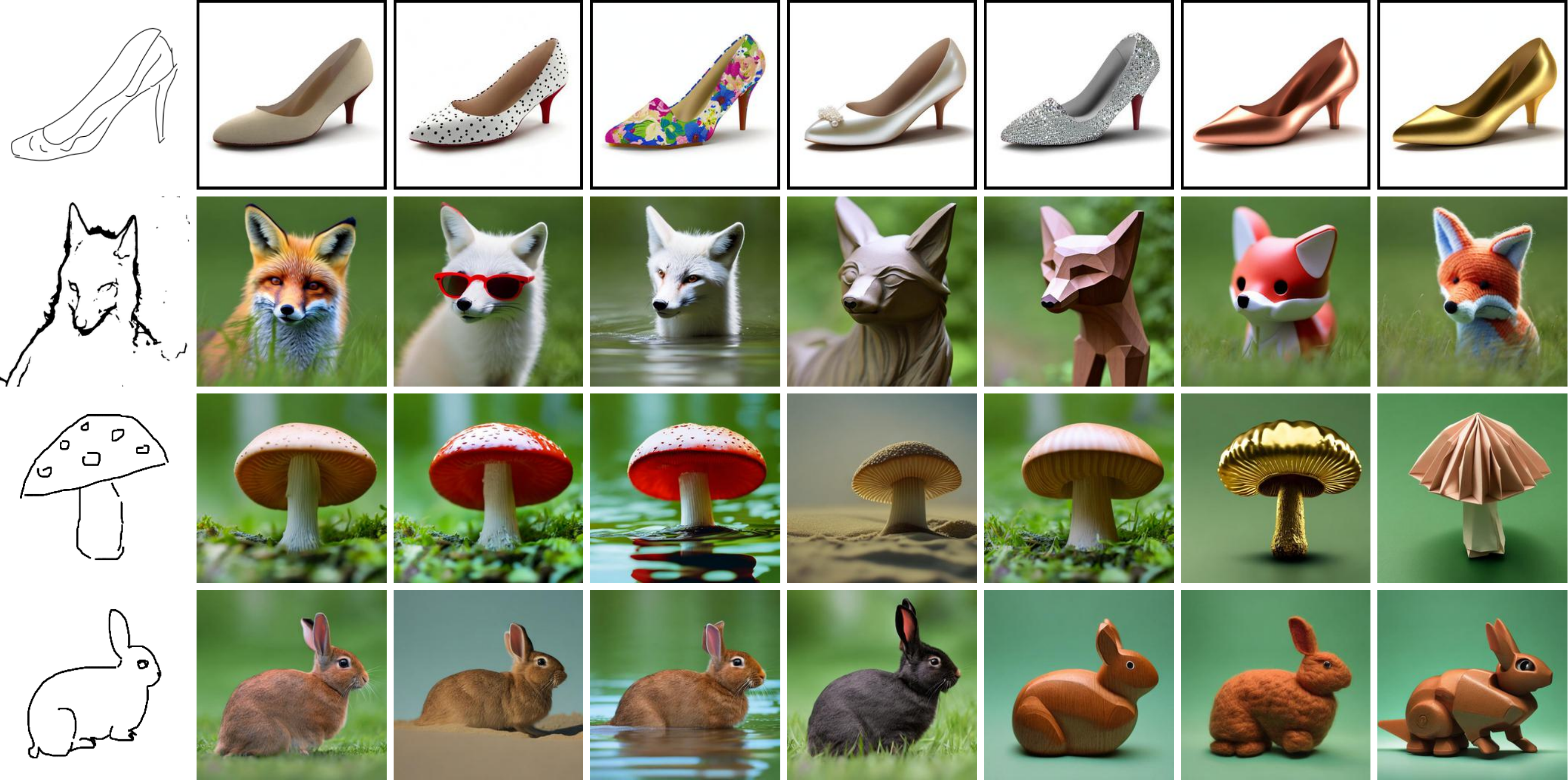}
  \caption{Visual attribute control in the proposed sketch-to-image translation method. Training and sampling can exclusively use freehand sketches (\textbf{first row}) or edge maps (\textbf{second row}). Alternatively, training can be performed on edge maps while sampling uses freehand sketches of unseen (\textbf{third row}) or known (\textbf{fourth row}) object classes.}
  \label{fig:visual_control}
  \vspace{-1.0em}
\end{figure}

\noindent
\textbf{Visual attribute control in the generated images:} One key advantage of the proposed method is the ability to control visual attributes in the generated images for general image editing and manipulation. As the architecture does not require retraining the LDM, we can use the pre-trained LDM as a learned prior for visual modifications alongside LCTN to impose structural constraints. Fig. \ref{fig:visual_control} shows a few examples where we render a specific object in multiple visual styles by providing different text prompts to the pre-trained LDM while keeping a consistent shape across different styles as intended in the input sketch.

%% file: sec/5_conclusions.tex
\section{Conclusions}
\label{sec:conclusions}

In this paper, we introduce a novel sketch-to-image translation technique that uses a learnable lightweight mapping network (LCTN) for latent code translation from sketch to image domain, followed by $k$ forward diffusion and $T$ backward denoising steps through a pre-trained text-to-image LDM. We show that by selecting an optimal value for $k \sim [1, T]$ near the upper threshold ($k \approx T$, $k < T$), it is possible to generate highly detailed photorealistic images that closely resemble the intended structures in the given sketches. Our experiments demonstrate that the proposed technique outperforms the existing methods in most visual and analytical comparisons across multiple datasets. Additionally, we show that the proposed method retains structural consistency across different visual styles, allowing photorealistic style manipulation in the generated images.